\documentclass{article}
\usepackage{graphicx}
\usepackage{multirow}
\usepackage{dcolumn}
\usepackage{cite}

\begin{document}

\title{Comparison of two semiclassical expansions for a family of PT-symmetric oscillators}
\author{Francisco M. Fern\'{a}ndez and Javier Garcia \\
INIFTA (UNLP, CCT La Plata-CONICET), Divisi\'on Qu\'imica
Te\'orica,\\ Blvd. 113 S/N,  Sucursal 4, Casilla de Correo 16,
1900 La Plata, Argentina} \maketitle

\begin{abstract}
We show that a recently developed semiclassical expansion for the
eigenvalues of PT-symmetric oscillators of the form $V(x)=(ix)^{2N+1}+bix$
does not agree with an earlier WKB expression for $V(x)=-(ix)^{2N+1}$ the
case $b=0$. The reason is due to the choice of different paths in the
complex plane for the calculation of the WKB integrals. We compare the
Stokes and anti-Stokes lines that apply to each case for the quintic
oscillator and derive a general WKB expression that contains the two earlier
ones.
\end{abstract}

\section{Introduction}

In a recent paper Nanayakkara and Mathanaranjan\cite{NM13} obtained accurate
high-energy expansions for the solutions to the Schr\"{o}dinger equation
with polynomial potentials of odd degree. By means of an alternative
approach, the asymptotic-energy expansion (AEE) that is claimed to be
simpler than the standard WKB method, the authors derived analytic
expressions which allowed them to obtain accurate eigenvalues for several
test examples. They compared their approximate eigenvalues with accurate
ones provided by both the diagonalization method and numerical integration.
The agreement between the AEE and numerical results is remarkable and
increases with the quantum number as expected.

The first set of test examples chosen by Nanayakkara and Mathanaranjan\cite
{NM13} is given by $V(x)=(ix)^{2N+1}+bix$. It is surprising that they did
not try to compare their results for $b=0$ with those of Bender and Boettcher%
\cite{BB98}. Such comparison would have been interesting because the latter
authors stated that the diagonalization method is unsuitable for the
PT-symmetric oscillators $V(x)=-(ix)^{K}$ when $K\geq 4$. The reason for
this failure is that the Stokes wedges that lie in the lower half of the
complex plane do not contain the real $x$ axis. It is therefore striking at
first sight that the AEE results of Nanayakkara and Mathanaranjan\cite{NM13}
agree so accurately with the ones provided by the diagonalization method.

Recently, we studied the eigenvalues of some non-Hermitian operators by
means of the complex-rotation diagonalization and Riccati-Pad\'{e} methods%
\cite{FG13}. We verified that the diagonalization method is unable to yield
the eigenvalues of the PT-symmetric oscillators for $K\geq 4$ and that it
produces instead some real eigenvalues related to the resonances discussed
more rigorously by Alvarez\cite{A88,A95} many years before.

It is clear from the results just discussed that the semiclassical
approaches proposed by Nanayakkara and Mathanaranjan\cite{NM13}
and Bender and Boettcher\cite{BB98} are not equivalent. The
purpose of this paper is to compare them and make clear the
discrepancy.

\section{Comparison of the WKB approaches}

\label{sec:comparison}

Nanayakkara and Mathanaranjan\cite{NM13} considered non-Hermitian
oscillators
\begin{equation}
H=p^{2}+V(x),
\end{equation}
with polynomial potentials of the form
\begin{equation}
V(x)=(ix)^{2N+1}+P(x),
\end{equation}
where $P(x)$ is a polynomial function of degree smaller than $2N+1$. For
concreteness we focus present discussion on the first case studied by those
authors: $P(x)=bix$.

The authors calculated the integrals that appear in the AEE along a contour
that encloses two of the $2N+1$ branch points. They stated that those
particular branch points lie inside the Stokes wedges which are necessary
for defining the ``above non-Hermitian problem correctly as an eigenvalue
problem'' and made reference to Shin's paper\cite{S05}. The latter author
studied non-Hermitian oscillators with potentials $V(z)=-(iz)^{m}+P(z)$,
where $P(z)$ is a polynomial function of degree smaller than $m$ and the
boundary conditions are such that the eigenfunctions tend to zero
exponentially as $|z|\rightarrow \infty $ along the rays
\begin{equation}
\arg (z)=-\frac{\pi }{2}\pm \frac{2\pi }{m+2}.  \label{eq:arg(z)}
\end{equation}
This condition is exactly the one required by Bender and Boettcher and,
consequently, the WKB expressions of Bender and Boettcher and Shin are
exactly the same. Therefore, it is intriguing how Nanayakkara and
Mathanaranjan based on the wedges for a slightly different problem (see the
sign of the leading term) derived an expression that obviously yields
different results but still seems to be sound. We analyze this question in
detail in what follows.

To begin with, we compare the leading term of the AEE when $b=0$:
\begin{equation}
E_{n}^{NM}=\left[ \frac{\sqrt{\pi }(2N+3)\Gamma \left( \frac{1}{2}+\frac{1}{%
2N+1}\right) }{2\cos \left( \frac{\pi }{4N+2}\right) \Gamma \left( \frac{1}{%
2N+1}\right) }\left( n+\frac{1}{2}\right) \right] ^{\frac{4N+2}{2N+3}},
\label{eq:E(NM)}
\end{equation}
and the WKB expression adapted to the present problem

\begin{equation}
E_{n}^{BB}=\left[ \frac{\sqrt{\pi }\Gamma {\left( \frac{3}{2}+\frac{1}{2N+1}%
\right) }}{\sin {\left( \frac{\pi }{2N+1}\right) }\Gamma {\left( 1+\frac{1}{%
2N+1}\right) }}\left( n+\frac{1}{2}\right) \right] ^{\frac{4N+2}{2N+3}},
\label{eq:E(BB)}
\end{equation}
where NM and BB stand for Nanayakkara and Mathanaranjan and Bender and
Boettcher, respectively. Before proceeding any further, it is worth noting
that the asymptotic eigenvalue $\lambda _{j}$ discussed by Shin\cite{S05}
(see also the references therein) agrees with the latter expression when $%
j=n+1$ and $m=2N+1$. Straightforward calculation shows that $%
E_{n}^{BB}>E_{n}^{NM}$ for all $N>1$ and that the discrepancy increases with
$N$ as shown by $E_{n}^{BB}/E_{n}^{NM}=1,\,1.988629015,\,3.523156867$ for $%
N=1,2,3$, respectively. We appreciate that both approaches agree only in the
case that the Stokes wedges chosen by Bender and Boettcher and Shin contain
the real $x$ axis ($N=1$).

Another noticeable difference is that Bender and Boettcher chose all the the
turning points on the lower half of the complex plane, whereas those of
Nanayakkara and Mathanaranjan appear alternatively in the upper and lower
half of the complex plane for $N=1,2,\ldots $.

We can obtain accurate eigenvalues of the PT-symmetric oscillators $%
V(x)=-(ix)^{2N+1}$ by means of the Riccati-Pad\'{e} method (RPM) applied
recently to this kind of problems\cite{FG13} (and references therein). Table~%
\ref{tab:En(x5)} shows the first eigenvalues of the PT-symmetric oscillator $%
V(x)=-(ix)^{5}$ calculated by the RPM and the WKB approach of Bender and
Boettcher. Table~\ref{tab:En(x5)b} shows the first eigenvalues of the
oscillator $V(x)=(ix)^{5}$ calculated by means of the diagonalization
method, numerical integration (see below) and the leading term of the AEE%
\cite{NM13}. We appreciate that the results of both tables do not agree as
expected from the discussion above.

\begin{figure}[]
~\bigskip\bigskip
\par
\begin{center}
\includegraphics[width=8cm]{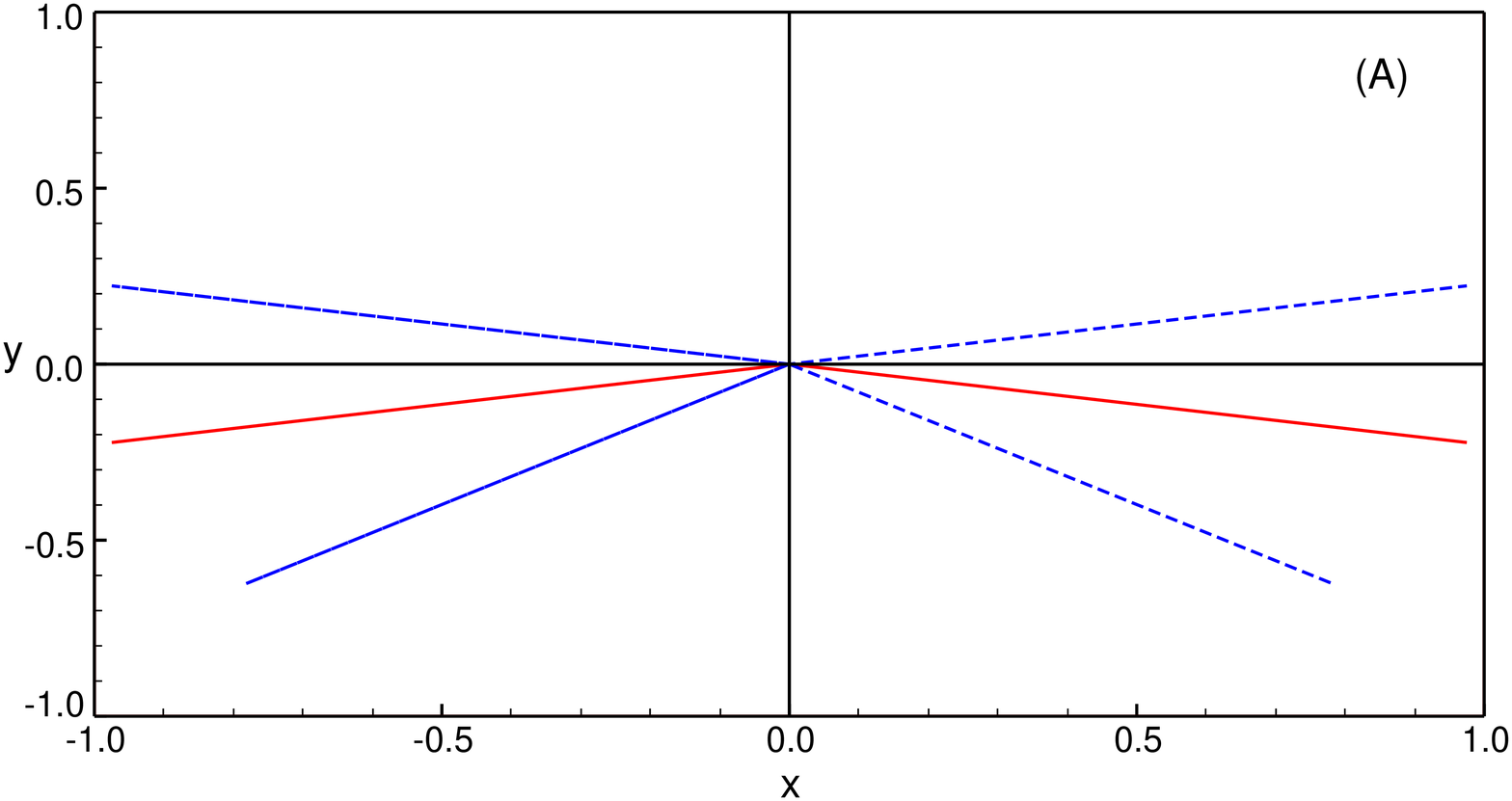} \includegraphics[width=8cm]{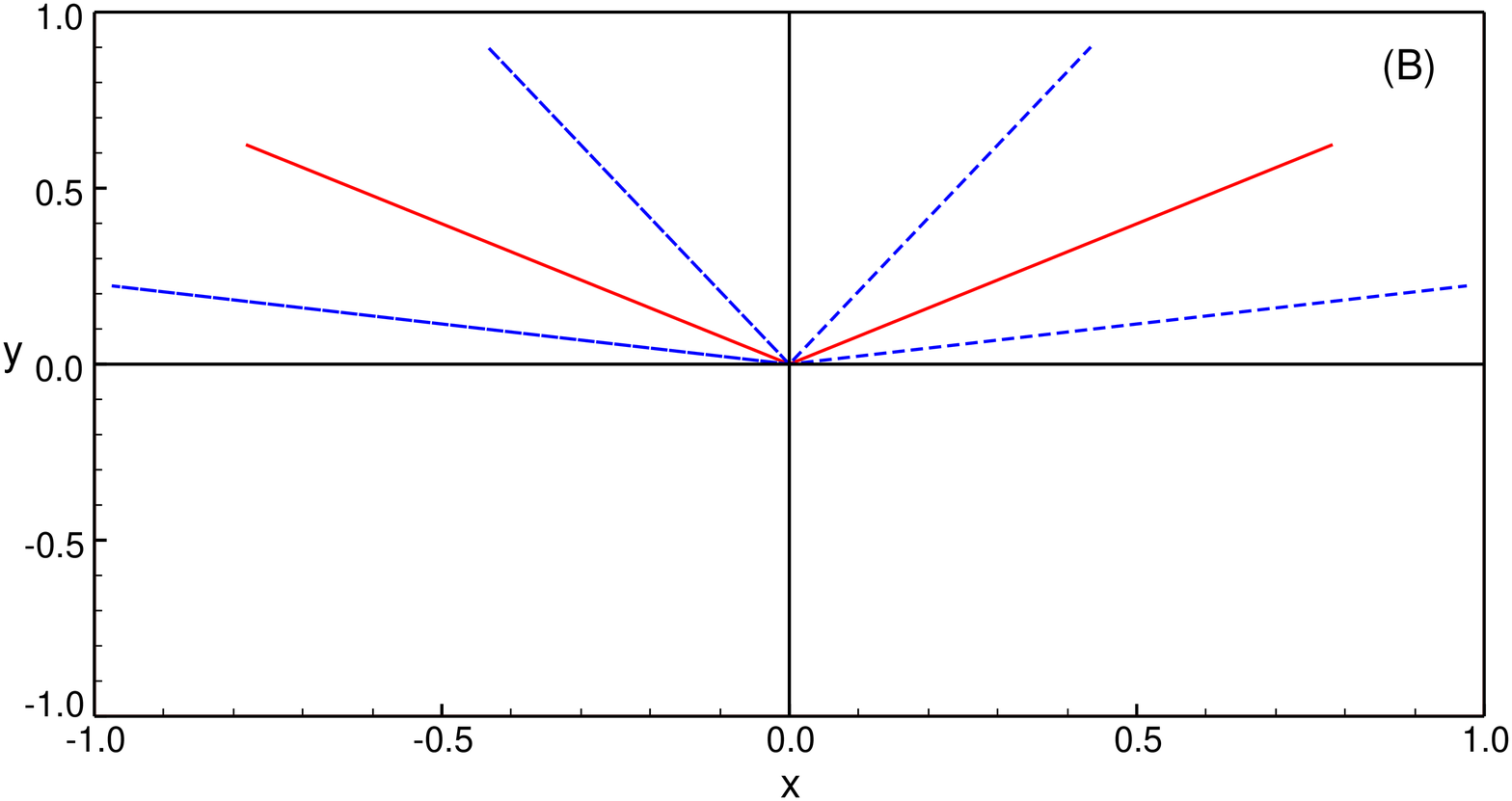}
\par
\bigskip
\end{center}
\caption{Some Stokes (dashed, blue) and anti-Stokes (solid, red) lines for
the PT-symmetric oscillator $V(x)=(ix)^5$. The lines are located at (A): $%
\frac{\pi}{14},\frac{13\pi}{14},\frac{15\pi}{14},\frac{17\pi}{14},\frac{25\pi%
}{14},\frac{27\pi}{14}$, (B): $\frac{\pi}{14},\frac{3\pi}{14},\frac{5\pi}{14}%
,\frac{9\pi}{14},\frac{11\pi}{14},\frac{13\pi}{14}$ }
\label{fig:WNM}
\end{figure}

\begin{figure}[]
~\bigskip\bigskip
\par
\begin{center}
\includegraphics[width=8cm]{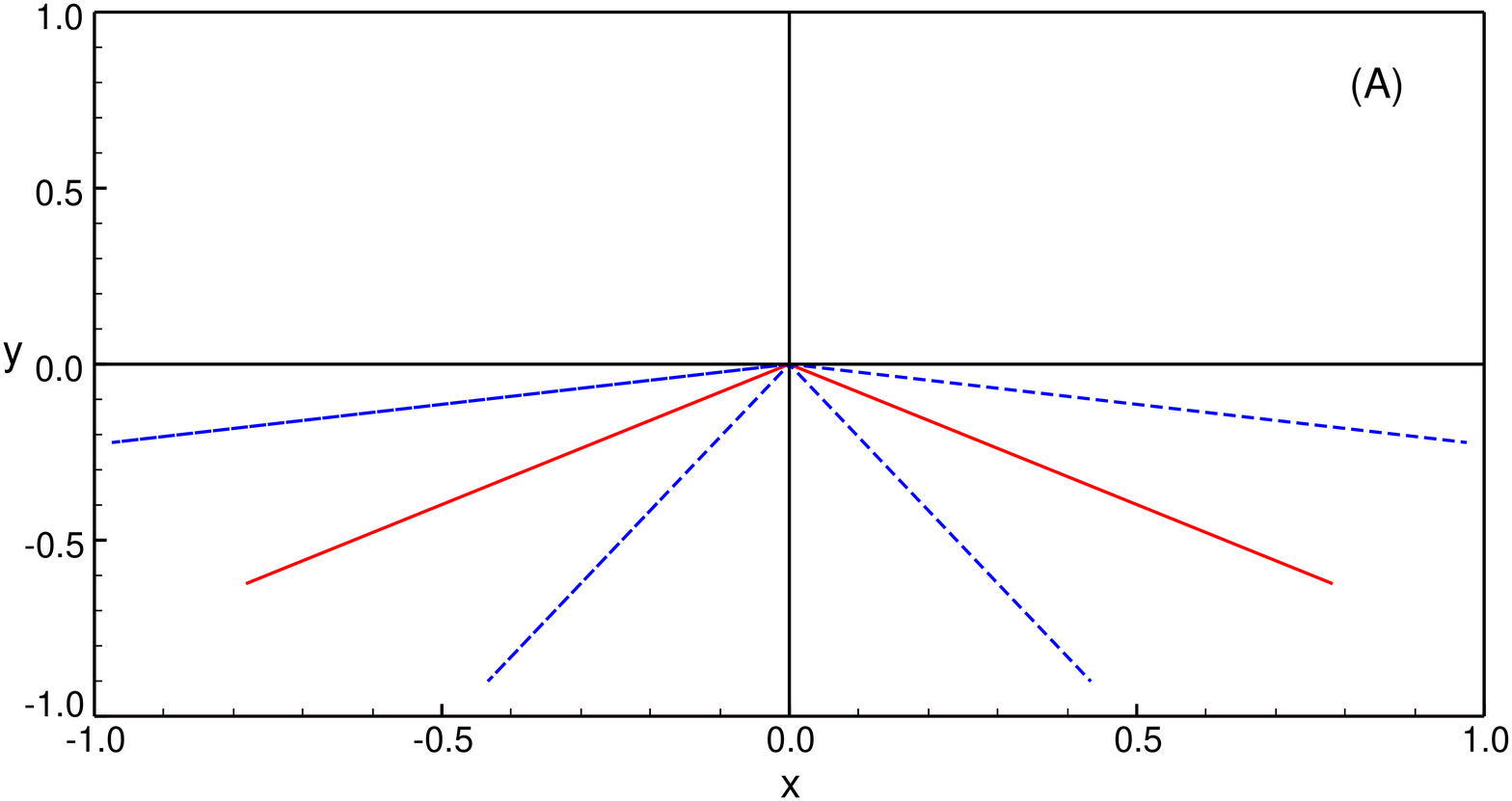} \includegraphics[width=8cm]{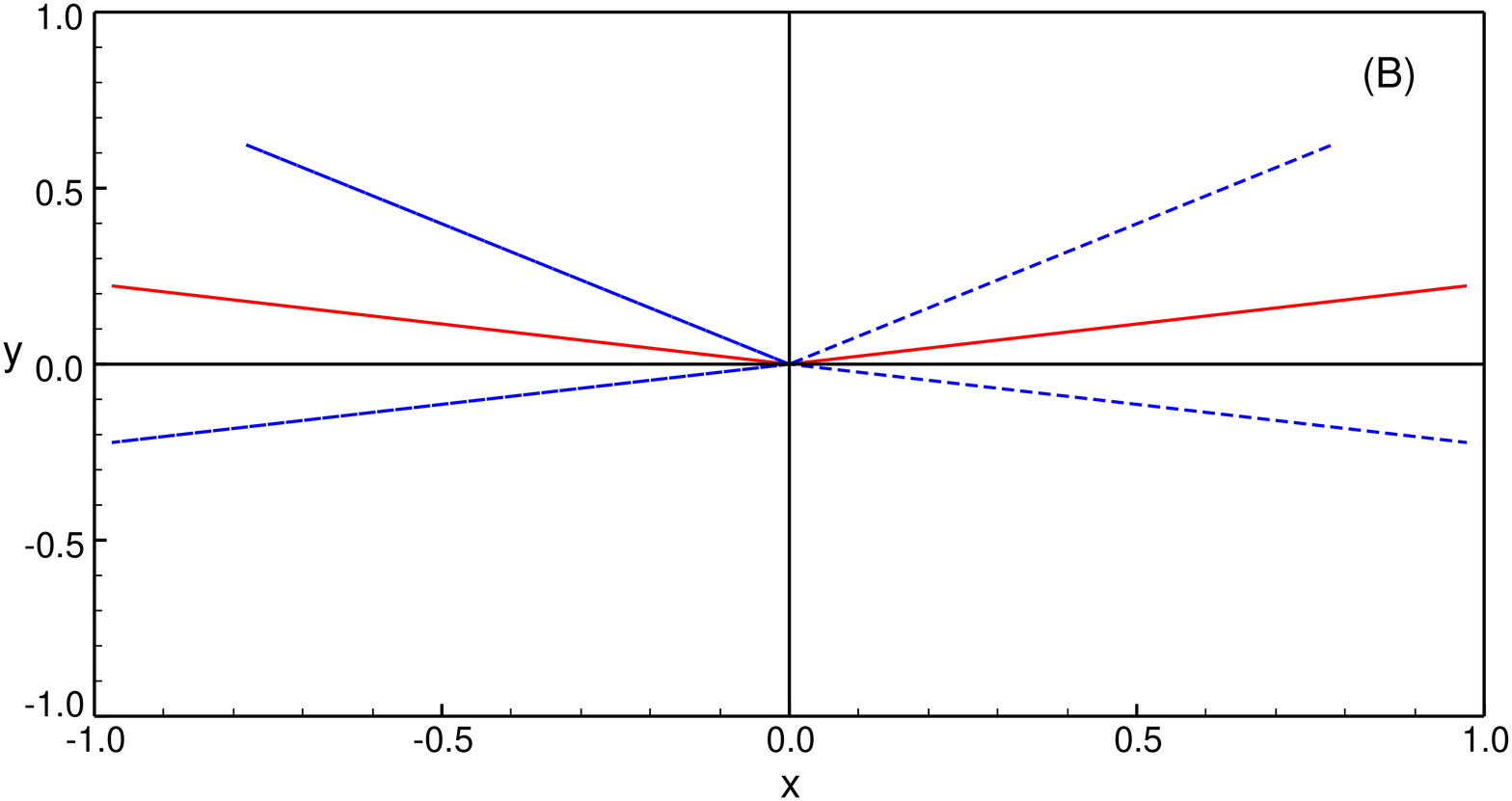}
\par
\bigskip
\end{center}
\caption{Some Stokes (dashed, blue) and anti-Stokes (solid, red) lines for
the PT-symmetric oscillator $V(x)=-(ix)^5$. The lines are located at (A): $%
\frac{15\pi}{14},\frac{17\pi}{14},\frac{19\pi}{14},\frac{23\pi}{14},\frac{%
25\pi}{14},\frac{27\pi}{14}$, (B): $\frac{\pi}{14},\frac{3\pi}{14},\frac{%
11\pi}{14},\frac{13\pi}{14},\frac{15\pi}{14},\frac{27\pi}{14}$ }
\label{fig:WBB}
\end{figure}

In order to understand the discrepancy between the AEE and WKB we carried
out numerical integrations of the Schr\"{o}dinger equation with $%
V(x)=(ix)^{5}$ along different anti-Stokes lines. Our results are consistent
with the AEE ones when we choose the anti-Stokes lines shown in figure~\ref
{fig:WNM} (A). In this case the Stokes wedges contain the real axis and
these results agree with those coming from the diagonalization method as
indicated above. We have recently shown that the optimal angle in the
complex-rotation method is the one that converts a non-Hermitian oscillator
in either a Hermitian or a PT-symmetric one\cite{FG13}. In the present case,
since we are dealing with PT-symmetric oscillators, the rotation angle is
expected to be exactly zero. Therefore, the straightforward diagonalization
with a real basis set (for example, harmonic oscillator eigenfunctions)
should give reasonable results as already shown in Table~\ref{tab:En(x5)b}.
On the other hand, Bender and Boettcher chose the Stokes wedges and
anti-Stokes lines shown in figure~\ref{fig:WBB} (A). In this case the Stokes
wedges do not contain the real axis and the eigenvalues cannot be obtained
by diagonalization. In order to obtain similar results for the potential $%
V(x)=(ix)^{5}$ one should choose the lines shown in figure~\ref{fig:WNM} (B)
in which case the Stokes wedges do not contain the real axis. From the
Schr\"{o}dinger equation with $V(x)=-(ix)^{5}$ we can also obtain the
results of Nanayakkara and Mathanaranjan if we choose the wedges indicated
in ~\ref{fig:WBB} (B).

\section{General WKB formula}

\label{sec:Gen_WKB}

In this section we will derive a general WKB expression that contains both
Bender and Boettcher's and Nanayakkara and Mathanaranjan's equations. To
this end we write the potential as
\begin{equation}
V(x)=x^{2M}(ix)^{\epsilon },
\end{equation}
where $M=1,2,\ldots $ and $\epsilon $ is chosen conveniently. The leading
term of the WKB energy is given by
\begin{equation}
\left( n+\frac{1}{2}\right) \pi =\int_{x_{-}}^{x_{+}}\sqrt{%
E-x^{2M}(ix)^{\epsilon }}dx,
\end{equation}
where the turning points $x_{\pm }$ are two roots of
\begin{equation}
E-x_{j}^{2M}(ix_{j})^{\epsilon }=0,
\end{equation}
that are given by
\begin{equation}
x_{j}=e^{-\frac{i\pi }{2}\left( \frac{\epsilon -4j}{2M+\epsilon }\right) }E^{%
\frac{1}{2M+\epsilon }}.
\end{equation}
Obviously, if $x_{j}$ is a root then $-x_{j}^{*}$ is also a root; therefore,
if we choose $x_{+}=x_{0}$ and $x_{-}=-x_{0}^{*}$ then we obtain
\begin{equation}
E_{n}\sim \left[ \frac{\sqrt{\pi }\Gamma \left( \frac{3}{2}+\frac{1}{%
2M+\epsilon }\right) \left( n+\frac{1}{2}\right) }{\sin \left( \frac{\pi M}{%
2M+\epsilon }\right) \Gamma \left( 1+\frac{1}{2M+\epsilon }\right) }\right]
^{\frac{4M+2\epsilon }{2M+\epsilon +2}}.  \label{eq:EJ}
\end{equation}

If $M=1$ and $\epsilon =2N-1$ we have $V(x)=x^{2}(ix)^{2N-1}=-(ix)^{2N+1}$
that is exactly the PT-symmetric potential chosen by Bender and Boettcher
for the particular case of odd oscillators. It is clear that for those
values of $M$ and $\epsilon $ the formula (\ref{eq:EJ}) becomes (\ref
{eq:E(BB)}). Besides, the corresponding turning points are also in complete
agreement.

In order to obtain the equation of Nanayakkara and Mathanaranjan we have to
proceed more carefully. If we choose $M=2L$ and $\epsilon =\pm 1$ we have $%
V(x)=x^{4L}(ix)^{\epsilon }=(ix)^{4L+\epsilon }$, where $L$ and $\epsilon $
are determined by the relation $4L+\epsilon =2N+1$. Alternatively, we may
set $M=N$ and $\epsilon =1$ so that equation (\ref{eq:EJ}) becomes (\ref
{eq:E(NM)}). It seems surprising at first sight that we obtain the correct
eigenvalues when the potential parameters lead to $%
V(x)=x^{2N}(ix)=(-1)^{N}(ix)^{2N+1}$. The reason is that in this case all
the turning points are on the lower half of the complex plane. In the former
case the form of the potential is kept fixed while the turning points shift
from one half of the plane to the other as $N$ changes by unity (exactly as
in Nanayakkara and Mathanaranjan' approach). On the other hand, in the
latter case the turning points are kept on the lower half of the complex
plane while the sign of the potential changes as $N$ changes. Both
strategies lead to the same eigenvalues and are embodied in the general
expression (\ref{eq:EJ}).

\section{Conclusions}

\label{sec:conclusions}

The purpose of this paper is to compare the AEE proposed by
Nanayakkara and Mathanaranjan\cite{NM13} and an earlier WKB
approach derived Bender and Boettcher\cite{BB98} whose expression
was already confirmed by Shin\cite{S05} (see also references
therein). We have shown that those approaches are not equivalent
and yield completely different results, except for $N=1$. The
reason lies on the choice of the Stokes wedges within which the
eigenfunction vanishes exponentially. While the Stokes wedges
chosen by Bender and Boettcher do not contain the real axis when
$N>1$ those chosen by Nanayakkara and Mathanaranjan contain it for
all $N$. For this reason the results of the latter authors agree
so accurately with the eigenvalues given by the diagonalization
method. Both semiclassical eigenvalues agree exactly for $N=1$
because the corresponding Stokes wedges contain the real axis\cite
{BB98}. Under such condition the diagonalization method with a
real basis set is known to be exactly equivalent to the
complex-rotation method because the optimal rotation angle is
zero\cite{FG13}.

Nanayakkara and Mathanaranjan\cite{NM13} also compared their AEE
eigenvalues with accurate results provided by numerical
integration of the differential equations. Unfortunately, they did
not give any detail that enables one to understand the nature of
their results. We think that present paper makes this point clear
by showing explicitly the Stokes and anti-Stokes lines that define
each of the alternative eigenvalue problems. Bender and
Boettcher\cite {BB98} also carried out a numerical integration of
the Schr\"{o}dinger equation which confirmed the accuracy of their
WKB eigenvalues. They found that convergence is most rapid when
one integrates along the anti-Stokes lines that they showed
explicitly in their paper toghether with the corresponding Stokes
lines.

Along with the choice of the Stokes wedges there is also the
question of the choice of the turning points for the calculation
of the WKB integrals. Whereas Bender and Boettcher chose all the
turning points in the lower half of the complex plane, those of
Nanayakkara and Mathanaranjan appear alternatively either in the
upper or lower half. From the comparison of both approaches we
have been able to derive the WKB expression (\ref{eq:EJ}) that
contains the equations derived by both Bender and Boettcher and
Nanayakkara and Mathanaranjan. It is undoubtedly the most
important point of present paper.

\begin{table}[]
\caption{Eigenvalues of the anharmonic oscillator $V(x)=(ix)^5$ calculated
by means of the Riccati-Pad\'e method (RPM) and the WKB expression of Bender
and Boettcher\protect\cite{BB98} ($E_n^{BB}$) }
\label{tab:En(x5)}
\begin{center}
\par
\begin{tabular}{|D{.}{.}{2}|D{.}{.}{41}|D{.}{.}{10}|}
\hline \multicolumn{1}{|c}{$n$} & \multicolumn{1}{|c|}{RPM} &
\multicolumn{1}{c|}{$E_n^{BB}$}
\\
\hline
0 & 1.9082645781707777079714407742647568531562     & 1.771244715 \\
1 & 8.58722083620722180027956616257834275867345   & 8.509035978  \\
2 & 17.710809011731145002460444521074221024       & 17.65253759  \\
3 & 28.595103311735974787298524540082589714       & 28.54706617    \\

\hline
\end{tabular}
\end{center}
\end{table}

\begin{table}[]
\caption{Eigenvalues of the anharmonic oscillator $V(x)=(ix)^5$ calculated
by means of diagonalization (DM), numerical integration along the
anti-Stokes lines (NI), and the leading term of the AEE\protect\cite{NM13}}
\label{tab:En(x5)b}
\begin{center}
\par
\begin{tabular}{|D{.}{.}{2}|D{.}{.}{14}|D{.}{.}{10}|D{.}{.}{10}|}
\hline \multicolumn{1}{|c}{$n$} & \multicolumn{1}{|c|}{DM} &
\multicolumn{1}{c|}{NI}
& \multicolumn{1}{c|}{$E_n^{NM}$}\\
\hline
0 & 1.16477040794341  & 1.164771    & 0.8906863480\\
1 & 4.36378436771211  & 4.363785    & 4.278845331 \\
2 & 8.95516699824067  & 8.955167    & 8.876737420 \\
3 & 14.4177548302741  & 14.417755   & 14.35514917   \\
4 &  20.6101375100489 &   20.610138 & 20.55551587 \\
5 &  27.4284077210062 &   27.428408 & 27.37969662 \\
6 &  34.8037156407346 &   34.803715 & 34.75941365 \\
7 &  42.6845638108818 &   42.684564 & 42.64372812   \\
8 &  51.030837828189  &   51.030837 & 50.99281286 \\
9 &  59.81014759020   &   59.810150 & 59.77445901 \\
10&  68.9956534721    &   68.995644 & 68.96194510 \\

\hline
\end{tabular}
\end{center}
\end{table}

\begin{thebibliography}{9}
\bibitem{NM13}  Nanayakkara A and Mathanaranjan T 2013 \textit{Phys. Scr.}
\textbf{88} 055004.

\bibitem{BB98}  Bender C M and Boettcher S 1998 \textit{Phys. Rev. Lett.}
\textbf{80} 5243.

\bibitem{FG13}  Fern\'{a}ndez F M and Garcia J 2013 \textit{J. Phys. A}
\textbf{46} 195301. arXiv:1301.1676 [math-ph]

\bibitem{A88}  Alvarez G 1988 \textit{Phys. Rev. A} \textbf{37} 4079.

\bibitem{A95}  Alvarez G 1995 \textit{J. Phys. A} \textbf{28} 4589.

\bibitem{S05}  Shin K C 2005 \textit{J. Phys. A} \textbf{38} 6147.
\end{thebibliography}
\end{document}